\documentclass[a4paper]{llncs}

\usepackage{dsfont,stmaryrd,amsmath,booktabs}

\usepackage{amssymb}
\usepackage{graphicx}

\usepackage{url}
\urldef{\mailsa}\path|{christophe.guyeux, qianxue.wang,|
\urldef{\mailsb}\path|jacques.bahi}@univ-fcomte.fr|   
\newcommand{\keywords}[1]{\par\addvspace\baselineskip
\noindent\keywordname\enspace\ignorespaces#1}

\begin{document}

\mainmatter  

\title{A Pseudo Random Numbers Generator Based on Chaotic Iterations.\\
Application to Watermarking}


%
%
\author{Christophe Guyeux\and Qianxue Wang\and Jacques M. Bahi}
%


\institute{University of Franche-Comte,
Computer Science Laboratory LIFC,\\
25030 Besan\c con Cedex, France\\
\mailsa\\
\mailsb\\
}

%
%

\maketitle

\begin{abstract}
In this paper, a new chaotic pseudo-random number generator (PRNG) is proposed. It combines the well-known ISAAC and XORshift generators with chaotic iterations. This PRNG possesses important properties of topological chaos and can successfully pass NIST and TestU01 batteries of tests. This makes our generator suitable for information security applications like cryptography. As an illustrative example, an application in the field of watermarking is presented.
\end{abstract}

%

\section{Introduction}
The extremely fast development of the Internet brings growing attention to information security issues. Among these issues, the conception of pseudo-random number generators (PRNGs) plays an important role. Secure PRNGs which can be easily implemented with simple software routines are desired. Due to the finiteness of the set of machine numbers, the sequences generated by numerous existing PRNGs are not actually random. For example, the use of stringent batteries of tests allows us to determine whether these sequences are predictable.
Chaos theory plays an active role in the improvement of the quality of PRNGs \cite{Cecen2009},~\cite{PO2004}. The advantage of using chaos in this field lies in its disordered behavior and its unpredictability.

This paper extends the study initiated in~\cite{guyeux09} and~\cite{wang2009}. In~\cite{guyeux09}, it is proven that chaotic iterations (CIs), a suitable tool for fast computing iterative algorithms, satisfy the topological chaotic property, as it is defined by Devaney~\cite{Dev89}. In~\cite{wang2009}, the chaotic behavior of CIs is exploited in order to obtain an unpredictable behavior for a new PRNG. This generator is based on chaotic iterations and depends on two other input sequences. These two sequences are generated by two logistic maps. Our generator has successfully passed the NIST (National Institute of Standards and Technology of the U.S. Government) battery of tests. However it appeared that it is a slow generator and it can't pass TestU01 because of the input logistic maps. Moreover this logistic map has revealed serious security lacks, which make it use inadequate for cryptographic applications~\cite{Arroyo08}. That is why, in this paper, we intend to develop a new fast PRNG. It will pass TestU01, widely considered as the most comprehensive and stringent battery of tests. This goal is achieved by using the ISAAC and XORshift maps in place of the two logistic maps. Chaotic properties, statistical tests and security analysis~\cite{ZHENG92008} allow us to consider that this generator has good pseudo-random characteristics and is capable to withstand attacks.

The rest of this paper is organized in the following way: in Section~\ref{Basic recalls}, some basic definitions concerning chaotic iterations and PRNGs are recalled. Then, the generator based on discrete chaotic iterations is presented in Section~\ref{The generation of pseudo-random sequence}. Section~\ref{Security analysis} is devoted to its security analysis. In Section~\ref{Testing a generator}, we show that the proposed PRNG passes the TestU01 statistical tests. In Section~\ref{An application example of the proposed PRNG} an application in the field of watermarking is proposed. The paper ends by a conclusion and some discussions about future work.
\section{Basic recalls}
\label{Basic recalls}
This section is devoted to basic notations and terminologies in the fields of chaotic iterations and PRNGs.
\subsection{Notations}
\begin{tabular}{@{}c@{}@{}l@{}}
$\llbracket 1;\mathsf{N} \rrbracket$ & $\rightarrow\{1,2,\hdots,N\}$ \\
$S^{n}$ & $\rightarrow$ the $n^{th}$ term of a sequence $S=(S^{1},S^{2},\hdots)$ \\
$v_{i}$ & $\rightarrow$ the $i^{th}$ component of a vector \\
&~~~~~~~$v=(v_{1},v_{2},\hdots, v_n)$\\
$f^{k}$ & $\rightarrow$ $k^{th}$ composition of a function $f$ \\

$\emph{strategy}$ & $\rightarrow$ a sequence which elements belong in $%
\llbracket 1;\mathsf{N} \rrbracket $ \\
$\mathbb{S}$ & $\rightarrow$ the set of all strategies \\
$\oplus$ & $\rightarrow$ bitwise exclusive or \\
$+$ & $\rightarrow$ the integer addition \\
$\ll \text{and} \gg$ & $\rightarrow$ the usual shift operators \\
\end{tabular}
\subsection{Chaotic iterations}
\label{subsection:Chaotic iterations}
\begin{definition}
The set $\mathds{B}$ denoting $\{0,1\}$, let $f:\mathds{B}^{\mathsf{N}%
}\longrightarrow \mathds{B}^{\mathsf{N}}$ be an ``iteration'' function and $S\in \mathbb{S}
$ be a chaotic strategy. Then, the so-called \emph{chaotic iterations} are defined by~\cite{Robert1986}
\begin{equation}
\begin{array}{l}
x^0\in \mathds{B}^{\mathsf{N}}, \\
\forall n\in \mathds{N}^{\ast },\forall i\in \llbracket1;\mathsf{N}\rrbracket%
,x_i^n=\left\{
\begin{array}{l}
x_i^{n-1} ~~~~~\text{if}~S^n\neq i \\
f(x^{n-1})_{S^n} ~\text{if}~S^n=i.\end{array} \right. \end{array}
\end{equation}
\end{definition}
In other words, at the $n^{th}$ iteration, only the $S^{n}-$th cell is
\textquotedblleft iterated\textquotedblright .

\subsection{Input sequences}
In \cite{wang2009}, we have designed a PRNG which has successfully passed the NIST tests suite. Unfortunately, this PRNG is too slow to pass the TestU01 battery of tests. Our ancient PRNG which is called CI(Logistic, Logistic) PRNG is based on chaotic iterations and uses logistic maps as input sequences. However, chaotic systems like logistic maps work in the real numbers domain, and therefore a transformation from real numbers into integers is needed. This process leads to a degradation of the chaotic behavior of the generator and a lot of time wasted during computations. Moreover, a recent study shows that the use of logistic map for cryptographic applications is inadequate and must be discouraged~\cite{Arroyo08}. Our purpose is then to design a new, faster, and more secure generator, which is able to pass the TestU01 battery of tests. This is achieved by using some faster PRNGs like ISAAC~\cite{Jenkins1996} and XORshift~\cite{Marsaglia2003} as input sequences.
\section{Design of CI(ISAAC,XORshift)}
\label{The generation of pseudo-random sequence}
\subsection{Chaotic iterations as PRNG}
The novel generator is designed by the following process. Let $\mathsf{N} \in \mathds{N}^*, \mathsf{N} \geqslant 2$. Some chaotic iterations are fulfilled to generate a sequence $\left(x^n\right)_{n\in\mathds{N}} \in \left(\mathds{B}^\mathsf{N}\right)^\mathds{N}$ of boolean vectors: the successive states of the iterated system. Some of these vectors are randomly extracted and their components constitute our pseudo-random bit flow.
Chaotic iterations are realized as follows. Initial state $x^0 \in \mathds{B}^\mathsf{N}$ is a boolean vector taken as a seed and chaotic strategy $\left(S^n\right)_{n\in\mathds{N}}\in \llbracket 1, \mathsf{N} \rrbracket^\mathds{N}$ is constructed with XORshift. Lastly, iterate function $f$ is the vectorial boolean negation
$$f_0:(x_1,...,x_\mathsf{N}) \in \mathds{B}^\mathsf{N} \longmapsto (\overline{x_1},...,\overline{x_\mathsf{N}}) \in \mathds{B}^\mathsf{N}.$$
To sum up, at each iteration only $S^i$-th component of state $X^n$ is updated, as follows
\begin{equation}
x_i^n = \left\{\begin{array}{ll}x_i^{n-1} & \text{if } i \neq S^i, \\ \\ \overline{x_i^{n-1}} & \text{if } i = S^i. \\\end{array}\right.
\end{equation}
Finally, let $\mathcal{M}$ be a finite subset of $\mathds{N}^*$. Some $x^n$ are selected by a sequence $m^n$ as the pseudo-random bit sequence of our generator. The sequence $(m^n)_{n \in \mathds{N}} \in \mathcal{M}^\mathds{N}$ is computed with ISAAC. So, the generator returns the following values: the components of $x^{m^0}$, followed by the components of $x^{m^0+m^1}$, followed by the components of $x^{m^0+m^1+m^2}$, \emph{etc.}
In other words, the generator returns the following bits:\newline
\begin{small}
$$x_1^{m_0}x_2^{m_0}x_3^{m_0}\hdots x_\mathsf{N}^{m_0}x_1^{m_0+m_1}x_2^{m_0+m_1}\hdots x_\mathsf{N}^{m_0+m_1} x_1^{m_0+m_1+m_2}x_2^{m_0+m_1+m_2}\hdots$$

\noindent or the following integers:$$x^{m_0}x^{m_0+m_1}x^{m_0+m_1+m_2}\hdots$$
\end{small}
The basic design procedure of the novel generator is summed up in Table~\ref{CI}.
The internal state is $x$, the output array is $r$. $a$ and $b$ are those computed by ISAAC and XORshift generators. Lastly, $c$ and $\mathsf{N}$ are constants and  $\mathcal{M}=\{$c, c+1$\}$ ($c\geqslant 3\mathsf{N}$ is recommended).

\begin{table}
\centering
\begin{tabular}{|l|}
\hline
~\textbf{Input}: the internal state $x$ \\
~~~~~~~~~~~($x$ is an array of $\mathsf{N}$ 1-bit words) \\
\hline
~\textbf{Output}: an array $r$ of $\mathsf{N}$ 1-bit words\\
\hline
~$a\leftarrow{ISAAC()}$;\\
~$m\leftarrow{a~mod~2+c}$\\
~\textbf{for} $i=0,\dots,m$ \textbf{do}\\
~~~~~~$b\leftarrow{XORshift()}$;\\
~~~~~~$S\leftarrow{b~mod~\mathsf{N}}$;\\
~~~~~~$x_S\leftarrow{ \overline{x_S}}$;\\
~\textbf{end for}\\
~$r\leftarrow{x}$;\\
~\textbf{return} $r$;\\
\hline
~\textbf{An arbitrary round of CI generator}~\\
\hline

\end{tabular}
\caption{CI algorithms}
\label{CI}
\end{table}

\subsection{Example}
In this example, $\mathsf{N} = 5$ and $\mathcal{M} = \{$4,5$\}$ are chosen for easy understanding.
The initial state of the system $x^0$ can be seeded by the decimal part of the current time. For example, the current time in seconds since the Epoch is 1237632934.484084, so $t = 484084$. $x^0 = t \text{ (mod 32)}$ in binary digits, then $x^0 = (1, 0, 1, 0, 0)$. $m$ and $S$ can now be computed from ISAAC and XORshift:
\begin{itemize}
\item $m$ = 4, 5, 4, 4, 4, 4, 5, 5, 5, 5, 4, 5, 4,...
\item $S$ = 2, 4, 2, 2, 5, 1, 1, 5, 5, 3, 2, 3, 3,...
\end{itemize}
Chaotic iterations are done with initial state $x^0$, vectorial logical negation $f_0$ and strategy $S$. The result is presented in Table 3. Let us recall that sequence $m$ gives the states $x^n$ to return: $x^4, x^{4+5}, x^{4+5+4}, \hdots$\newline
\begin{table*}[!t]
\caption{Application example}
\label{table application example}
\centering
\begin{tabular}{c|ccccc|cccccc|cccccc}
\hline\hline
$m:$ & & & 4 & & & & & 5 & & & & & & 4 & & & \\ \hline
$S$ & 2 & 4 & 2 & 2 & & 5 & 1 & 1 & 5 & 5 & & 3 & 2 & 3 & 3 & & \\ \hline
In this$x^{0}$ & & & & & $x^{4}$ & & & & & & $x^{9}$ & & & & & $x^{13}$ & \\
1 & & & & &
1 & & $\xrightarrow{1} 0$ & $\xrightarrow{1} 1$ & & &
1 & & & & &
1 & \\
0 & $\xrightarrow{2} 1$ & & $\xrightarrow{2} 0$ & $\xrightarrow{2} 1$ &
1 & & & & & &
1 & & $\xrightarrow{2} 0$ & & & 0 &\\
1 & & & & &
1 & & & & & &
1 & $\xrightarrow{3} 0$ & & $\xrightarrow{3} 1$ & $\xrightarrow{3} 0$ &
0 &\\
0 & & $\xrightarrow{4} 1$ & & &
1 & & & & & &
1 & & & & &
1 &\\
0 & & & & &
0 & $\xrightarrow{5} 1$ & & & $\xrightarrow{5} 0$ & $\xrightarrow{5} 1$ &
1 & & & & &
1 &\\
\hline\hline
\end{tabular}\\
\vspace{0.5cm}
Binary Output: $x_1^{0}x_2^{0}x_3^{0}x_4^{0}x_5^{0}x_1^{4}x_2^{4}x_3^{4}x_4^{4}x_5^{4}x_1^{9}x_2^{9}x_3^{9}x_4^{9}x_5^{9}x_1^{13}x_2^{13}... = 10100111101111110...$
Integer Output:
$x^{0},x^{0},x^{4},x^{6},x^{8}... = 20,30,31,19...$

\end{table*}
So, in this example, the generated binary digits are: 10100111101111110011... Or the integers are: 20, 30, 31, 19...
\subsection{Chaotic iterations and chaos}
Generally the success of a PRNG depends, to a large extent, on the following criteria: uniformity, independence, storage efficiency, and reproducibility. A chaotic sequence may have these good pseudo-random criteria and also other chaotic properties, such as: ergodicity, entropy, and expansivity. A chaotic sequence is extremely sensitive to the initial states. That is, even a minute difference in the initial state of the system can lead to enormous differences in the final state even over fairly small timescales. Therefore, chaotic sequence well fits the requirements of pseudo-random sequence. Contrary to ISAAC or XORshift, our generator possesses these chaotic properties.

However, despite a huge number of papers published in the field of chaos-based PRNGs, the impact of this research is rather marginal. This is due to the following reasons: almost all PRNG algorithms using chaos are based on dynamical systems defined on continuous sets (e.g., the set of real numbers). So these generators are usually slow, require considerably more storage spaces, and lose their chaotic properties during computations. These major problems restrict their use as generators~\cite{Kocarev2001}. Moreover, even if the algorithm obtained by the inclusion of chaotic maps is itself chaotic, the implementation of this algorithm on a machine can cause it lose its chaotic nature. This is due to the finite nature of the machine numbers set.\newline
In this paper we don't simply integrate chaotic maps hoping that the implemented algorithm remains chaotic. The PRNG algorithms we conceive are constituted by discrete chaotic iterations that we mathematically proved in \cite{guyeux09}, that produce topological chaos as defined by Devaney. In the same paper, we raised the question of their implementation, proving in doing so that it is possible to design a chaotic algorithm and a chaotic computer program.
In conclusion, the generator proposed in this paper does not inherit its chaotic properties from a continuous real chaotic map, but from discrete chaotic iterations defined in Section \ref{subsection:Chaotic iterations}. As quoted above, it has been proven in~\cite{guyeux09} that chaotic iterations behave as chaos, as it is defined by Devaney: they are regular, transitive and sensitive to initial conditions. This famous definition of a chaotic behavior for a dynamical system implies unpredictability, mixture, sensitivity and uniform repartition. This allows the conception of a new generation of chaotic PRNGs. Because only integers are manipulated in discrete chaotic iterations, the chaotic behavior of the system is preserved during computations, and these computations are fast.

\section{Security analysis}
\label{Security analysis}
In this section a security analysis of the proposed generator is given.
\subsection{Key space}
The PRNG proposed in this document is based on discrete chaotic iterations. It has an initial value $x^0\in \mathds{B}^{\mathsf{N}}$. Considering this set of initial values alone, the key space size is equal to $2^\mathsf{N}$. In addition, this PRNG combines digits of two other PRNGs: ISAAC and XORshift. Let $k_1$ and $k_2$ be the key spaces of ISAAC and XORshift. So the total key space size is close to $2^\mathsf{N}\cdot k_1\cdot k_2$. Finally, the impact of $\mathcal{M}$ must be taken into account. This leads to conclude that the key space size is large enough to withstand attacks.
\subsection{Key sensitivity}
This PRNG is highly sensitive to the initial conditions. To illustrate this property proved in~\cite{guyeux09}, several initial values are put into the chaotic system. Let $H$ be the number of differences between the sequences obtained in this way. Suppose $n$ is the length of these sequences. Then the variance ratio $P$, defined by $P = H / n$, is computed. The results are shown in Figure~1a ($x$ axis is sequence lengths, $y$ axis is variance ratio $P$). Variance ratios approach $0.50$, which indicates that the system is extremely sensitive to the initial conditions.
\begin{figure}
\centering
\begin{tabular}{cc}
\includegraphics[scale=0.4]{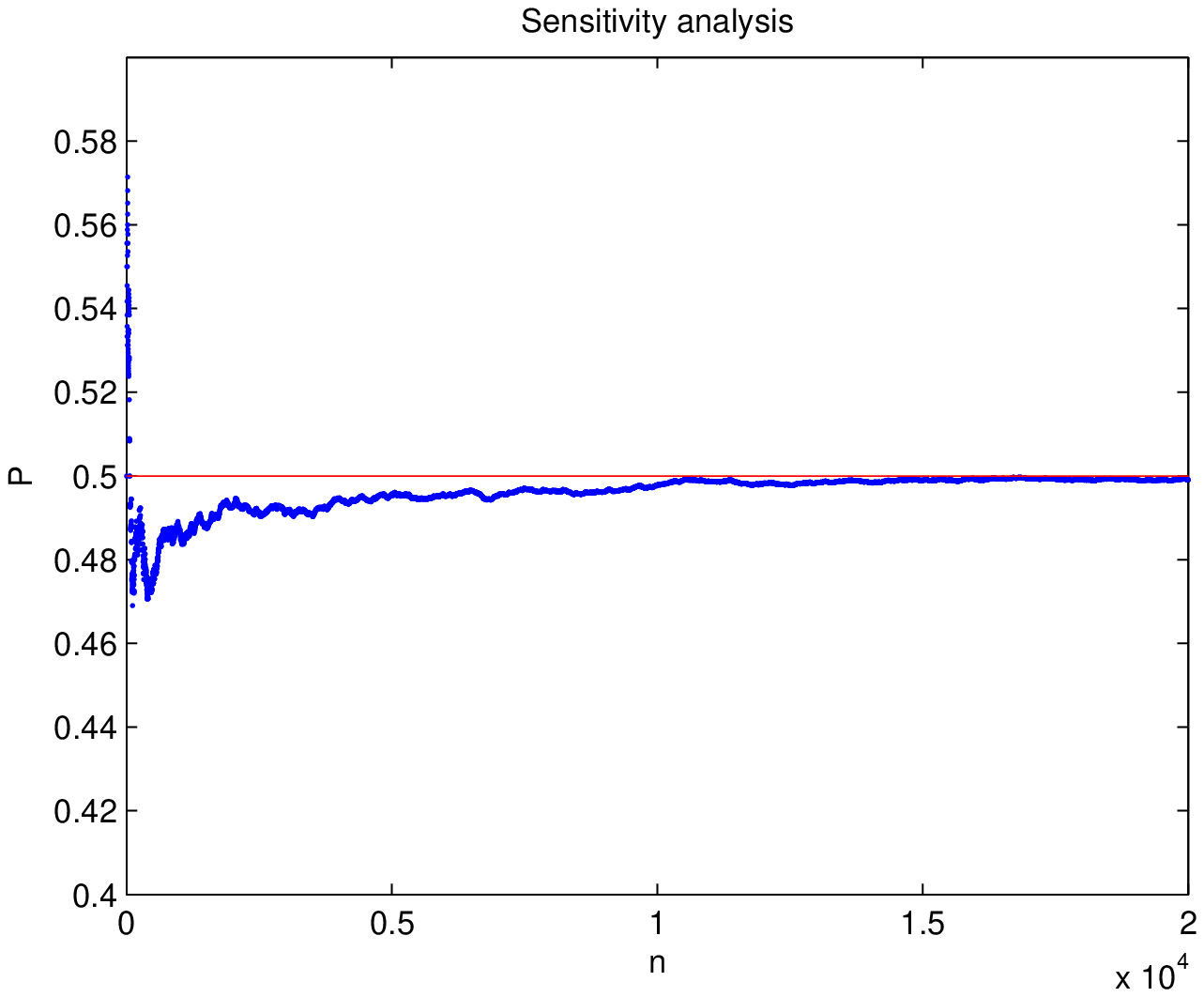} & \includegraphics[scale=0.4]{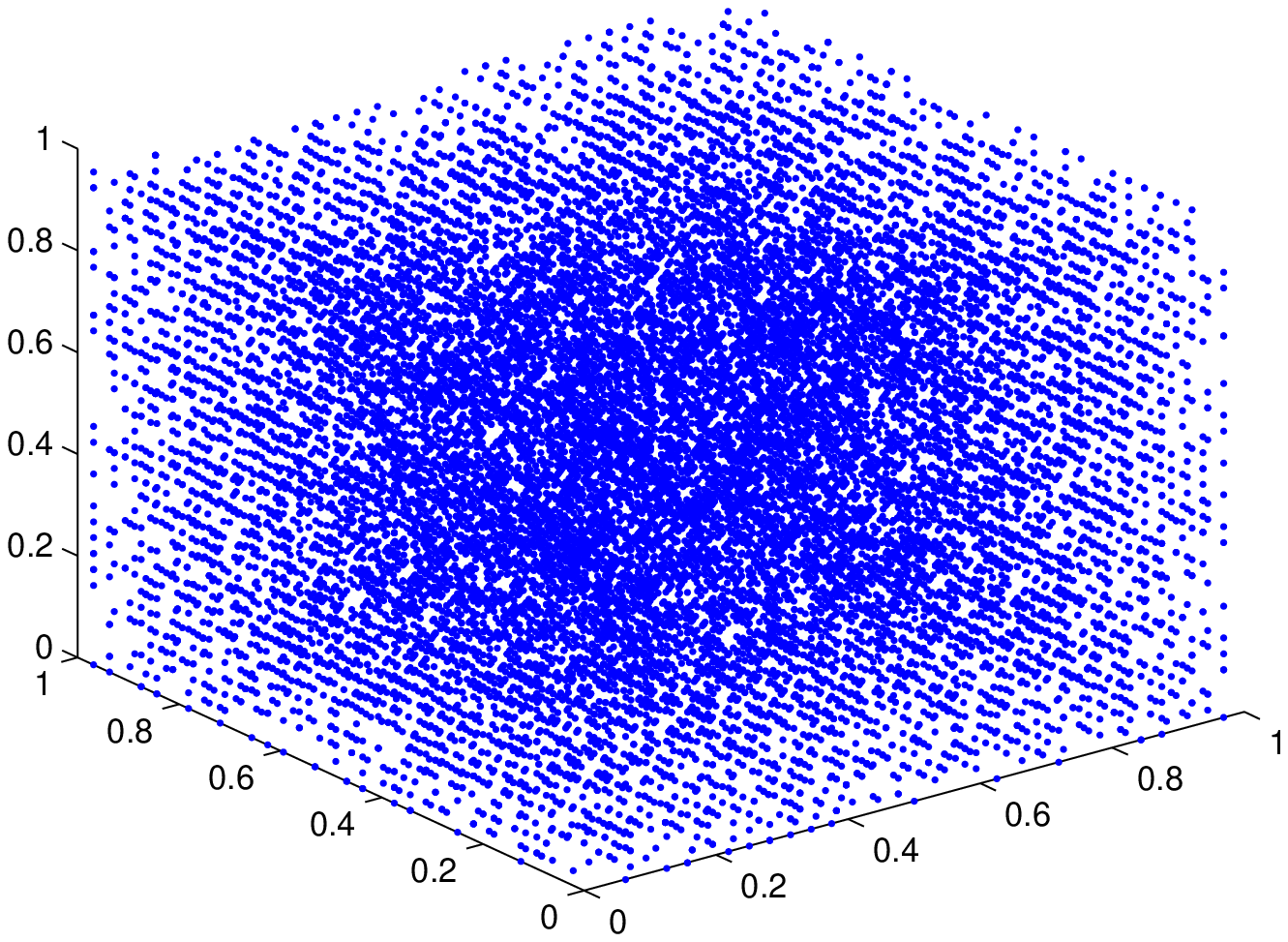} \\
\emph{a.} sensitivity & \emph{b.} Second order distribution \\
\end{tabular}
\DeclareGraphicsExtensions.
\label{SecurityAnalysis}
\caption{Security analysis}
\end{figure}

\subsection{Uniform distribution }
Figure~1b gives a 3D graphic representation of the distribution of a random sequence obtained by our generator. The point cloud presents a uniform distribution that tends to fill the complete 3D space, as expected for a random signal. To obtain this cloud, we have first changed the binary sequence to a $N$-bit integer sequence $x_1$, $x_2$, $x_3$, $x_4$... Then we have plot $\left(\frac{x_1}{2^N},\frac{x_2}{2^N},\frac{x_3}{2^N}\right), \left(\frac{x_2}{2^N},\frac{x_3}{2^N},\frac{x_4}{2^N}\right)$...

\section{TestU01 Statistical Test Results}
\label{Testing a generator}
In a previous section, we have shown that the proposed PRNG has strong chaotic properties, as Devaney's chaos. In particular, this generator is better than the well-known XORshift and ISAAC, in the topological point of view. In addition to being chaotic, we will show in this section that CI(ISAAC,XORshift) is better than XORshift, and at least as good as ISAAC~\cite{Wichmanna2006} in the statistical point of view. Indeed, similarly to ISAAC and contrary to XORshift, CI(ISAAC,XORshift) can pass the stringent Big Crush battery of tests included in TestU01. In addition, our generator achieves to pass all the batteries included in TestU01. To our best knowledge, this result has not been proven for ISAAC, and only one other generator is capable of doing this~\cite{Corsaro2009} 

\subsection{TestU01}

Indeed, the quality of a PRNG should be based on theoretical fundamentals but should also be tested
empirically. Various statistical tests
are available in the
literature that test a given sequence for some level of computational indistinguishability.
Major test suites for RNGs are TestU01~\cite{L'ecuyer2009}, the NIST suite~\cite{ANDREW2008}, and the
DieHARD suites~\cite{Marsaglia1996}. The DieHARD suites, which implement many classical RNG tests, have some drawbacks
and limitations. The National Institute of Standards and
Technology (NIST), in the United States, has implemented a test suite (16 tests) for RNGs. It
is geared mainly for
the testing and certification of RNGs used in cryptographic applications. TestU01
is extremely diverse in implementing classical tests,
cryptographic tests, new tests proposed in the literature, and original tests.
In fact, it encompasses most of the other test suites. The proposed PRNG has been tested using
TestU01 for its statistical pseudo randomness.
\subsection{Batteries of tests}
Table~\ref{TestU01 Statistical Test} lists seven batteries of tests in the TestU01 package. "Standard" parameter in this Table refers to the built-in parameters of the battery.
TestU01 suite implements 518 tests and reports $p-$values. If a $p-$value is within $[0.001,0.999]$, the associated test is a success. A $p-$value lying outside this boundary means that its test has failed. 

\begin{table}[!t]
\renewcommand{\arraystretch}{1.3}
\caption{TestU01 Statistical Test}
\label{TestU01 Statistical Test}
\centering
\begin{tabular}{c@{\quad}ccccc}\toprule
\textbf{Battery} &\textbf{ Parameters} & \textbf{Statistics} \\\midrule
Rabbit &$32\times10^9$ bits &40 \\
Alphabit &$32\times10^9$ bits &17 \\
Pseudo DieHARD &Standard &126 \\
FIPS\_140\_2 &Standard &16 \\
Small Crush &Standard &15 \\
Crush &Standard &144 \\
Big Crush &Standard &160 \\ \bottomrule
\end{tabular}
\end{table}

\subsection{Analysis}
\label{Analysis}
In a sound theoretical basis, a PRNG based on discrete chaotic iterations (ICs) is a composite generator which combines the features of two PRNGs. The first generator constitutes the initial condition of the chaotic dynamical system. The second generator randomly chooses which outputs of the chaotic system must be returned. The intention of this combination is to cumulate the effects of chaotic and random behaviors, to improve the statistical and security properties relative to each generator taken alone.

This PRNG based on discrete chaotic iterations may utilize any reasonable RNG as inputs. For demonstration purposes, XORshift and ISAAC are adopted here. The PRNG with these inputs can pass all of the performed tests.

\section{Application example in digital watermarking}
\label{An application example of the proposed PRNG}
In this section, an application example is given in the field of digital watermarking: a watermark is encrypted and embedded into a cover image using chaotic iterations and our PRNG. The carrier image is the famous Lena, which is a 256 grayscale image, and the watermark is the $64\times 64$ pixels binary image depicted in Fig.2d.
\begin{figure}
\centering
\begin{tabular}{lc}
\includegraphics[scale=0.25]{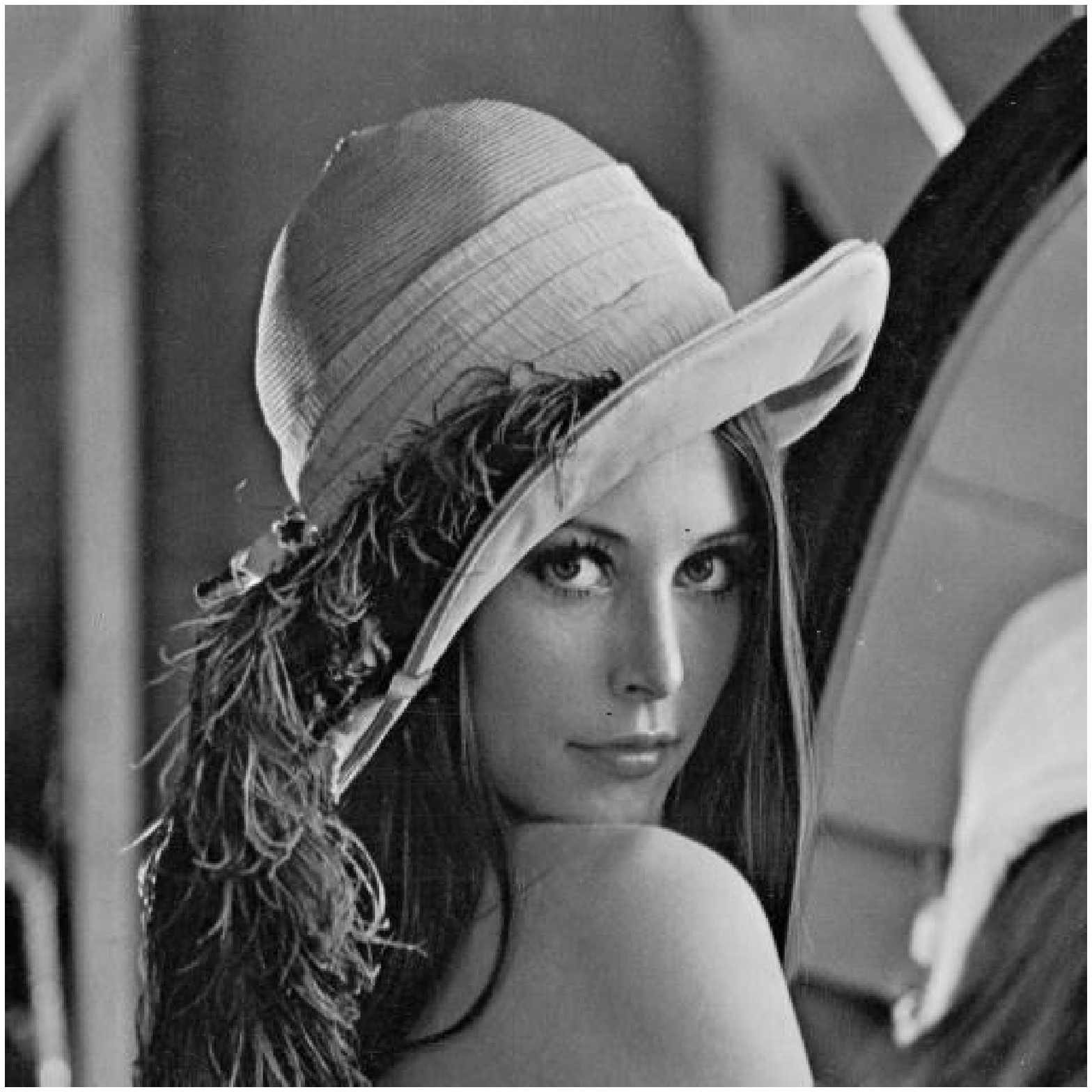} & ~~~~~~~~~~~~~~\includegraphics[scale=0.5]{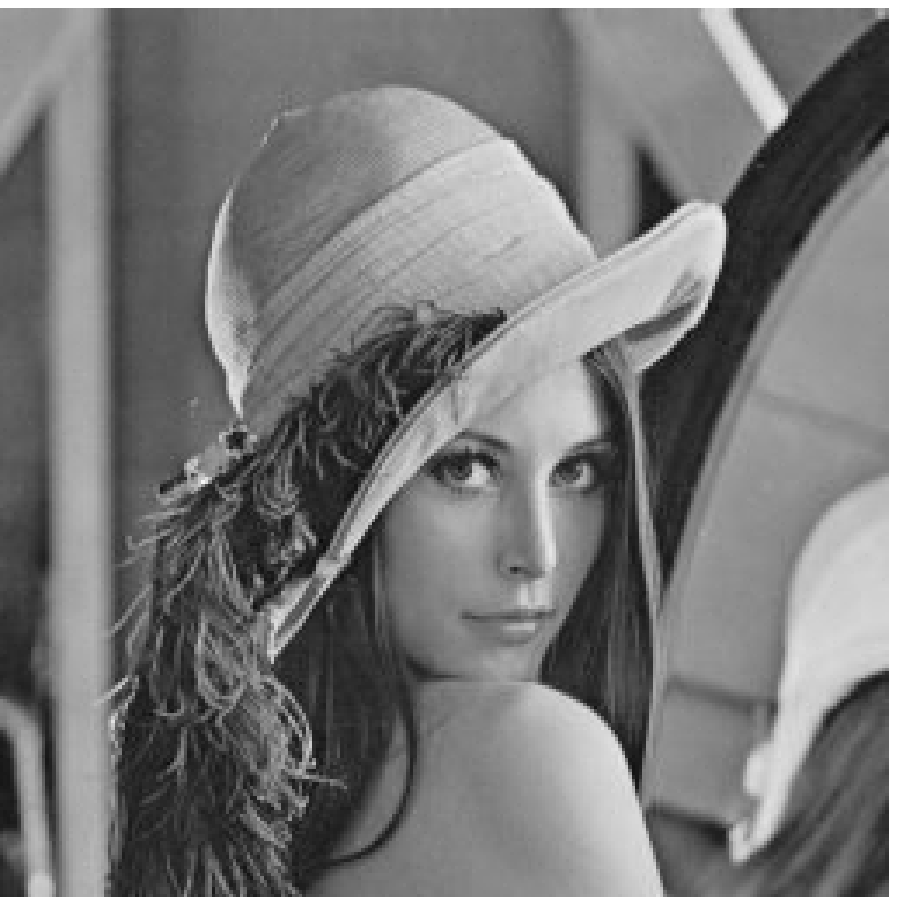}  \\
\emph{a.} Lena (\emph{scale} 0.5) & \emph{b.} Watermarked Lena \\
\end{tabular}
\begin{tabular}{lcr}
\includegraphics[scale=0.5]{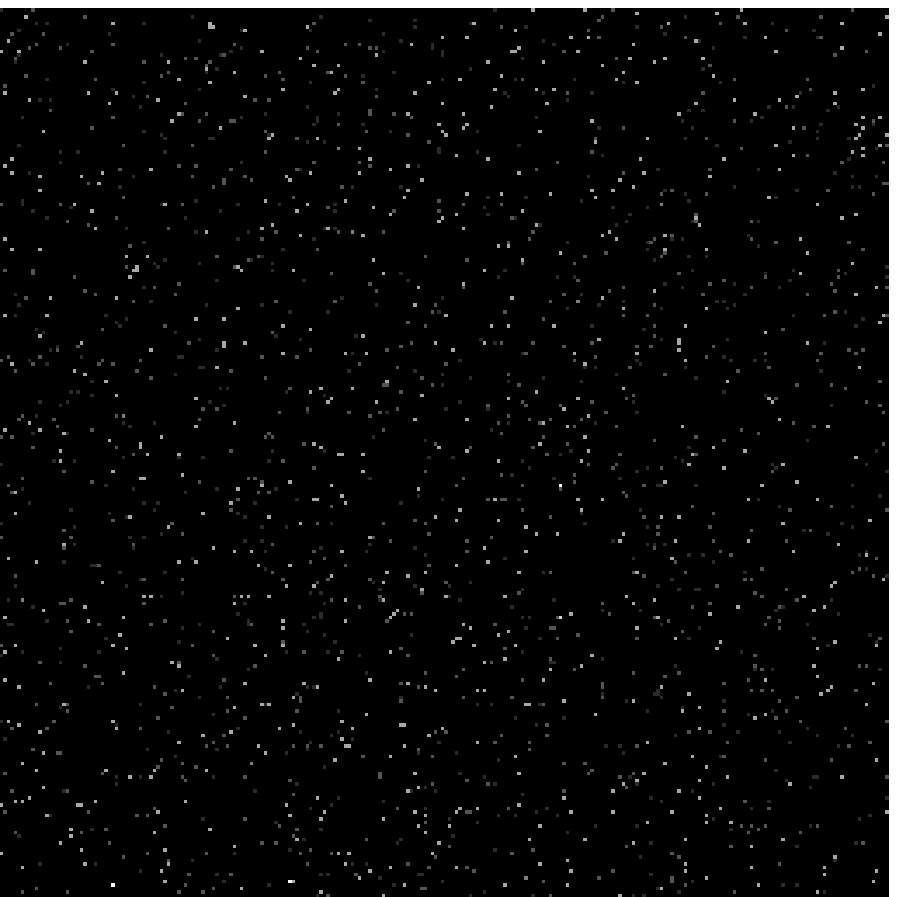}& ~~~~~~~~~~~~\includegraphics[scale=0.8]{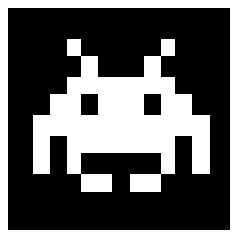} & ~~~~~~~~~~~~\includegraphics[scale=0.8]{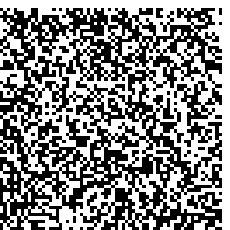}\\
 \emph{c.} Differences&\emph{d.} Watermark & \emph{e.} Encrypted  \\
&&watermark\\
\end{tabular}
\label{fig:Lenas}
\caption{Original and watermarked Lena}
\end{figure}
Let us encrypt the watermark by using chaotic iterations. The initial state
$x^{0}$ of the system is constituted by the watermark, considered as a boolean
vector. The iteration function is the vectorial logical negation $f_{0}$.
The PRNG presented previously is used to obtain a sequence of integers lower than 4096, which will constitute the chaotic strategy $(S^{k})_{k\in \mathds{N}}$. Thus, the encrypted watermark is the last boolean vector generated by the chaotic iterations. An example of such an encryption, with $5000$ iterations, is given in Fig.2e.

Let $L$ be the $256^3$ booleans vector constituted by the three last bits of each pixel of Lena. We define $U^k$ by $U^{0} = S^{0}$ and $U^{n+1} = S^{n+1}+2\times U^{n}+n ~ [mod ~ 256^3]$. The watermarked Lena $I_w$ is obtained from the original Lena $I_o$, the three last bits of which are replaced by the result of $64^2$ chaotic iterations with initial state $L$, and strategy $U^k$ (see Fig.2b). Spatial domain embedding has been chosen here for easy understanding, but this watermarking scheme can be adapted to frequency domain (for an example of its use in DWT domain, see~\cite{guyeux10ter}).
The extraction of the watermark can be obtained in the same way~\cite{guyeux10ter}. Remark that the map $\theta \mapsto 2\theta $ of the torus, which is the well-known dyadic transformation (an example of topological chaos \cite{Dev89}), has been chosen to make $(U^{k})_{k \leqslant 64^2}$ highly sensitive to the chaotic encryption strategy. 

The robustness of this data hiding scheme through geometric and frequency attacks has been studied in~\cite{guyeux10ter}. The chaos-security and stego-security are proven in~\cite{friot10}. 
The difference with the scheme presented in these papers is the way to generate strategies, \emph{i.e.}, the choice of the initial conditions for chaotic iterations, in the encryption and embedding stages. This improvement does not alter robustness and subspace-security. We have shown in this study that this replacement enhances the speed of the scheme. Moreover, it resolves a potential security lack related to the use of a logistic map~\cite{Arroyo08} when generating the strategies: this lack might be exploited by an attacker in Watermark-Only-Attack and Known-Message-Attack setups~\cite{Cayre2008}. Instead of logistic map, our PRNG has good statistical properties and can withstand such attacks. This claim will be deepened in future work.

\section{Conclusions and future work}
\label{Conclusions and Future Work}
In this paper, the PRNG proposed in \cite{wang2009} is improved. This is achieved by using the famous ISAAC and XORshift generators and by combining these components with chaotic iterations. Thus we obtain a faster generator which satisfies chaotic properties. In addition to passing the NIST tests suite, this new generator successfully passes all the stringent TestU01 battery of tests. The randomness and disorder generated by this algorithm has been evaluated. It offers a sufficient level of security for a whole range of applications in computer science. An application example in the field of data hiding is finally given. In future work, the comparison of different chaotic strategies will be explored and other iteration functions will be studied. Finally, other applications in computer science security field will be proposed, especially in cryptographic domains.
\bibliographystyle{plain}
\bibliography{Generating_good_chaotic_random_numbers.bib}
\end{document}